\documentclass[prd,aps]{revtex4}

\newcommand{\p}[4]{#1^{#3}_{#2,\,#4}}

\newcommand{\D}{\displaystyle}

\begin{document}
\title{Dirac-like approach for consistent discretizations
of classical constrained theories}

\author{Cayetano Di Bartolo$^{1}$, Rodolfo Gambini$^{2}$, 
Rafael A. Porto$^{3}$ and 
Jorge Pullin$^{4}$}
\affiliation {
1. Departamento de F\'{\i}sica,
Universidad Sim\'on Bol\'{\i}var,\\ Aptdo. 89000, Caracas 1080-A,
Venezuela.\\ 
2. Instituto de F\'{\i}sica, Facultad de Ciencias,
Igu\'a 4225, esq. Mataojo, Montevideo, Uruguay. \\ 
3. Department of Physics, Carnegie Mellon University, Pittsburgh,
PA 15213\\
4. Department of Physics and Astronomy, Louisiana State University,
Baton Rouge, LA 70803-4001}

\begin{abstract}
We analyze the canonical treatment of classical constrained mechanical
systems formulated with a discrete time. We prove that under very
general conditions, it is possible to introduce nonsingular canonical
transformations that preserve the constraint surface and the Poisson
or Dirac bracket structure. The conditions for the preservation of the
constraints are more stringent than in the continuous case and as a
consequence some of the continuum constraints become second class upon
discretization and need to be solved by fixing their associated
Lagrange multipliers. The gauge invariance of the discrete theory is
encoded in a set of arbitrary functions that appear in the generating
function of the evolution equations. The resulting scheme is general
enough to accommodate the treatment of field theories on the lattice. 
This paper attempts to clarify and put on sounder footing a
discretization technique that has already been used to treat a variety
of systems, including Yang--Mills theories, BF-theory and general
relativity on the lattice.
\end{abstract}

\maketitle

\section{Introduction}

We have recently introduced \cite{DiGaPu,GaPu} a technique for
treating the theories that arise when one discretizes (space)-time in
a constrained mechanical system or a continuum field theory. We have
shown that this technique works for Yang--Mills and BF theories and
implemented it for the gravitational case. Previous attempts to
studying systems with discrete time had concentrated mostly on systems
without constraints or with holonomic constraints only (for a review
with a comprehensive reference list see \cite{mawe}.)

The idea consists on starting from a discretized action, constructing
discrete Lagrange equations and introducing a symplectic structure in
the discrete space.  The evolution is implemented via canonical
transformations and the consistency of the discrete theory determines
in part the Lagrange multipliers. In some totally constrained
systems, like general relativity, the resulting discrete theories
are constraint-free since the constraints are solved for the
Lagrange multipliers. This makes the quantization of the discrete
theories considerably simpler than the continuum cases. This was
exploited to make progress in solving the problem of time in
quantum gravity \cite{greece,njp} and to implement the 
Page--Wootters relational time \cite{deco,njp} and show that a 
fundamental decoherence arises in quantum mechanics from quantum
gravity.

In this paper, we want to address in a more systematic way the issue
of the canonical formulation of discrete constrained systems. Up to
now, most of the analysis has been made on specific examples, and a
canonical analysis, a la Dirac\cite{Dirac,teit}, is still
lacking. In particular, the technique relied heavily on defining
a canonical transformation that was initially singular (and therefore
not a true canonical transformation) and showing that one could
eliminate variables and end up with a true canonical transformation.
Up to now this was shown in a case by case basis. This paper
determines the general conditions needed for the construction of 
a proper canonical transformation by following a close analogue of
the Dirac procedure, adapted to the discrete case. In particular,
we note that there are several ways to proceed that yield equivalent
results, but that may offer different advantages for particular
systems.

In section II we will lay out the framework of how to deal with
mechanical systems where time is discrete, including singular
and non-singular systems. In section III we will develop a
classification of constraints into first and second class suitable for
the discrete context. In section IV we work out a specific example
that exhibits the details of the formalism. In section V we discuss an
alternative formulation of the formalism and we end with conclusions
and discussion.

\section{Mechanics with discrete time}

We start by considering a continuum theory representing a mechanical
system. Its Lagrangian will be denoted by ${\hat L(q^a,\dot{q}^a)}$,
$a=1\ldots M$.  This setting is general enough to accommodate, for
instance, totally constrained systems. In such case $\dot{q}$ will be
the derivative of the canonical variables with respect to the
evolution parameter. It is also general enough to include the
systems that result from formulating on a discrete space-time lattice
a continuum field theory.

We discretize the evolution parameter in intervals (possibly varying
upon evolution) $t_{n+1}-t_n=\epsilon_n$ and we label the generalized
coordinates evaluated at $t_n$ as $q_n$. We define the discretized
Lagrangian as

\begin{equation} 
L(n,n+1) \equiv L(q^a_n,q^a_{n+1}) \equiv \epsilon_n{\hat
L}(q^a, \dot q^a)\label{disc} \end{equation}
where
\begin{equation} q=q_n \quad \hbox{and} \quad \dot q \equiv
{{q_{n+1}-q_n}\over \epsilon_n} \end{equation}

The action can then be written as
\begin{equation}
S=\sum_{n=1}^N L(n,n+1)
\end{equation}
If the continuum theory is invariant under
reparameterizations of the evolution parameter, one can show that the
information about the intervals $\epsilon_n$ may be absorbed in the
Lagrange multipliers. In the case of standard mechanical systems it is
simpler to use  an invariant interval $\epsilon_n=\epsilon$.  

The Lagrange equations of motion are obtained by requiring the
action to be stationary under variations of the configuration
variables $q^a$ fixed at the endpoints of the evolution
interval $n=0,n=N+1$,
\begin{equation}
{\partial L(n,n+1) \over \partial  q^a_{n}}+{\partial L(n-1,n)
\over \partial q^a_{n}}=0. \label{lagra}
\end{equation}

These equations define a unique evolution if the determinant
\begin{equation}
\left|{\partial^2 L(n,n+1) \over \partial  q^b_{n+1}\partial
 q^a_{n}}\right|\neq 0.\label{det0}
\end{equation}

We will refer to this case as the nonsingular case. When the
determinant vanishes, one will have to analyze the situation 
differently. Let us start with the nonsingular case.

\subsection{Nonsingular case}

In this case one can solve the Lagrange equations explicitly and the
$q_{n+1}$ are uniquely given as a function of $q_n$ and
$q_{n-1}$. This is the equivalent of the Hessian condition for the
non-singular Lagrangian theories in the continuum. The resulting
equations are ``second order'' in the sense that the $q_n$'s 
are determined provided one knows two previous time levels. One
can introduce a ``first order'' formulation by introducing 
canonically conjugate variables as is usually done when introducing
a Hamiltonian formulation in the continuum theories.

We introduce the following definition of variables that we will
later show end up being canonically conjugate momenta of the 
configuration variables,
\begin{eqnarray}
 p^a_{n+1} &\equiv& {\partial L(n,n+1) \over \partial  q^a_{n+1}}
\label{1}\\
 p^a_{n} &\equiv& {\partial L(n-1,n) \over \partial  q^a_{n}}=-
{\partial L(n,n+1) \over \partial  q^a_{n}}\label{2}
\end{eqnarray}

Where we have used Eq. (\ref{lagra}). The equations (\ref{1}) and
(\ref{2}) define a canonical transformation for the variables
$q_n,p_n$ to $q_{n+1},p_{n+1}$ with a the type 1
generating function $F_1= -L( q^a_n, q^a_n+1)$ provided that
condition (\ref{det0}) is fulfilled. Notice that the evolution 
scheme is implicit, one can use the bottom equation (since we
are in the non-singular case) to give an expression for 
$q_{n+1}$ in terms of $q_n,p_n$, which in turn can be substituted
in the top equation to get an equation for $p_{n+1}$ purely in
terms of $q_n,p_n$.

It should be noted that there are several other possible choices,
when going from the set of equations (\ref{1},\ref{2}) to 
an explicit evolution scheme. 
For example, one can choose to do things in a way that yields a closer
analogy with the standard Hamiltonian description in the continuum by
introducing type two canonical transformations.
To do this, we choose to invert equation (\ref{1}) for $
q^a_{n+1}\equiv  q^a_{n+1}( q^b_n, p^b_{n+1})$, which is 
possible only if 
\begin{equation}
\left|{\partial^2 L(n,n+1) \over \partial  q^a_{n+1}\partial
 q^b_{n+1}}\right|\neq 0.\label{det}
\end{equation}

We can  now introduce a Legendre transform and define
\begin{equation}
F_2(q_n,p_{n+1})\equiv \sum_a q^a_{n+1} p^a_{n+1} -L(q_n,q_{n+1}).
\end{equation}

{}From here it is immediate to obtain
\begin{eqnarray}
q^b_{n+1}&=& {\partial F_2 \over \partial p^b_{n+1}}\\ {\partial
F_2\over \partial q^b_n}&=&{-\partial L(n,n+1)\over \partial
q^b_n} =p^b_n,\label{3}
\end{eqnarray}
where we have used the Lagrange equation in order to obtain the
last equality. We easily recognize from here that $F_2$ behaves as
a type two generating function of the canonical transformation
connecting level $n$ with level $n+1$.

We may define now a sort of ``type 2'' Hamiltonian (in the 
sense that it depends on $p_{n+1}$ and $q_n$,
$H_2(p_{n+1},q_{n})$ given by,
\begin{equation}\label{Ham2}
H_2(q_{n},p_{n+1}) \equiv  \sum_a p^a_{n+1} (q^a_{n+1} - q^a_n) -
L(q_n,q_{n+1}) =F_2(q_{n},p_{n+1}) -  \sum_a p^a_{n+1} q^a_{n}\,.
\end{equation}
which leads to the discrete Hamilton equations,
\begin{eqnarray}
q^b_{n+1} &=& q^b_{n} +{\partial H_2 \over \partial p^b_{n+1}}\\
p^b_{n} &=& p^b_{n+1} +{\partial H_2 \over \partial
q^b_{n}}.\label{ec1}
\end{eqnarray}

It should be noted that although this formulation has a degree
of analogy with the traditional Hamiltonian formulation, there
are significant differences due to the fact that the conjugate
variables live at different time slices. It would not be 
possible therefore to use this formulation to attempt 
to construct a Schr\"odinger equation starting from the 
above Hamiltonian.

Provided that the canonical map defined by $F_2$ is invertible we
end up with a discrete evolution implemented by a canonical
transformation. It can be easily seen by using the Legendre
transform that $F_2$ will be invertible if and only if
\begin{equation}
\left|{\partial^2 F_2(n,n+1) \over \partial
 q^a_{n} p^b_{n+1}}\right| =-\left|{\partial^2 L(n,n+1) \over
\partial  q^a_{n} q^b_{n+1}}\right|\times \left|{\partial^2 L(n,n+1) \over
\partial  q^a_{n+1} q^b_{n+1}}\right|^{-1}\neq 0. \label{det2}
\end{equation}

Thus, in order to have canonical transformations generated by type
two functions, the Hessian condition in the continuum time
mechanics leads to two independent conditions in the discrete
theory given by Eqs. (\ref{det0}) and (\ref{det}). Notice however
that (\ref{det}) is not necessary for introducing a symplectic
structure.

It is clear that when one builds a canonical discrete theory there are
four possibilities depending on which pair of variables one chooses to
construct the generating functional of the canonical transformation,
either $q_n,q_{n+1}$, $q_n,p_{n+1}$, $p_n,q_{n+1}$, $p_n,p_{n+1}$. In
this subsection we considered only two, but the others can be easily
generalized from the discussion here.

\subsection{The singular case}

Let us consider as before the generic discrete Lagrangian $L(n,n+1)=L(
q^a_n, q^a_{n+1})$ with $a=1\ldots M$. It leads to the equations we 
already discussed,
\begin{eqnarray}
p^a_{n+1} &=& {\partial L(n,n+1) \over \partial
q^a_{n+1}}\label{P1}
\\
p^a_n  &=& -{\partial L(n,n+1) \over \partial
 q^a_{n}}\label{P2}
\end{eqnarray}

If $\D\left|{\partial^2 L(n,n+1) \over \partial q^a_{n+1}\partial
 q^b_{n}}\right|$ vanishes and its rank is $K$, the system is
singular and has $M-K$ constraints of the form
\begin{equation}
\Phi_A( q^a_n, p^a_n)=0
\end{equation}
that result from Eq(\ref{P2}), and $M-K$ constraints
\begin{equation}
\Psi_A( q^a_{n+1}, p^a_{n+1})=0
\end{equation}
resulting from Eq(\ref{P1}). The evolution of the
configuration variables from level $n$ to $n+1$ is given by
solving for $q^a_{n+1}$ the Eqs. (\ref{P2}).  As the system is
singular the evolution depends of $(M-K)$ arbitrary functions 
$V^A$
\begin{equation}
q^a_{n+1}=f^a( q^b_n, p^b_n,V^A).\label{evo1}
\end{equation}

We shall follow closely the standard Dirac canonical procedure
of continuum mechanics. In this case, the analysis of a constraint
system goes trough two steps. The first step consists in the
definition of a set of evolution equations that weakly preserve
the constraints and the Poisson symplectic structure. To do that
one defines the total Hamiltonian $H_T=H_0+V^\alpha\phi_\alpha$
where $\phi_\alpha$ are the primary constraints, and the
$V^\alpha$ are partially determined in order to preserve all the
constraints of the system. Even though some of the $V's$ may be
arbitrary functions, once they are specified, the evolution
generated by $H_T$ preserves the Poisson brackets and the
dynamical evolution is consistent at the classical level with the
constraint structure. The second step is only required to quantize
the system and consists in the identification of the first and
second class constraints and the introduction of the Dirac
brackets that enforce strongly the second class constraints.

As we shall see, the same procedure may be followed in the
discrete case. The main difference is the implementation of the
canonical transformation that is not generated by a Hamiltonian
but by a canonical transformation of type 2, 3, or 4.

Let us start by completing the evolution equations. We need to add
to Eq(\ref{evo1}) an equation for $p_{n+1}$
\begin{equation}
p^a_{n+1}=\left. {\partial L(q_n,q_{n+1}) \over
\partial  q^a_{n+1}}\right|_{ q^a_{n+1}=f^a( q^b_n,
p^b_n,V^A)} .\label{evo2}
\end{equation}

\noindent We now impose the preservation of the constraints

\begin{equation}
\Phi_A( q^a_{n+1}, p^a_{n+1}) = \Phi_A \left( {f^a}_n,{\partial
L(q_n,f^a) \over
\partial  q^a_{n+1}} \right)=0 \label{cons1}\,.
\end{equation}
Furthermore, we need to impose the $n+1$ level
constraints at level $n$, $\Psi_A( q^a_{n}, p^a_n)=0$ and impose
the consistency conditions
\begin{equation}
\Psi_A \left( {f^a}_n,{\partial L(q_n,f^a) \over
\partial  q^a_{n+1}} \right) =0\,.\label{cons2}
\end{equation}

Three different cases may occur:
\begin{itemize}
\item[a)] Eqs.(\ref{cons1},\ref{cons2})
vanish automatically, and therefore we are not led to new
conditions.
\item[b)] They lead to inconsistencies, and the dynamical
system is inconsistent.
\item[c)] New secondary constraints $C( q^a_n,p^a_n)$
appear or/and some of the arbitrary functions $V^A$ are
determined, that is $V^A=V^A( q^a_n,p^a_n,v^\alpha)$ with
$\alpha=1 \ldots R \le (N-K)$, and $v^\alpha$ arbitrary functions.
The process is repeated until consistency is achieved. That is,
until the consistency conditions are automatically satisfied
without further constraints and conditions for $V$.
\end{itemize}

Substituting $V^A$ in (\ref{evo1}) and (\ref{evo2}) we get the
evolution equations that preserve all the constraints: primary,
secondary, tertiary, and so on,
\begin{equation}
 q^a_{n+1}=f^a( q^b_n,p^b_n,V^A(q,p,v))={\tilde
f}^a( q^b_n,p^b_n,v^\alpha) \label{fun1}
\end{equation}
\begin{equation}
p^a_{n+1}=g^a( q^b_n, p^b_n,v^\alpha)\label{fun2}
\end{equation}

Initial values need to be restricted by the $n=0$ level constraints,
\begin{equation}
\Phi_A(q_0^a, p_0^a)= \Psi_A(q_0^a, p_0^a)=0\,.\label{Vinculos(0)}
\end{equation}

In order to have a complete analogy with the continuum  case, we
still need to analyze under what conditions this evolution also
preserves the Poisson bracket structure. As in the continuum case
we assume that the arbitrary functions $v$ have been fixed. Three
different cases may arise depending on if one chooses to implement
things in terms of a canonical transformation of type 2, 3 or 4:

{\em Case I)} The equation(\ref{fun1}) is invertible for $q^a_n$:
that is, $\D\left|{\partial q^a_{n+1} \over \partial q^b_n}\right|
\neq 0$ and therefore one can write
\begin{equation}
q^a_{n}=h^a( q^b_{n+1}, p^b_n).\label{inv1}
\end{equation}

Notice that, under these hypotheses, there are no
pseudo-constraints of the form $G(q^a_{n+1},p^a_n)=0$. We call
these pseudo-constraints because they involve variables at
different instants of time. 

We may define a type 3 generating functions of canonical
transformations,
\begin{equation}
F_3( q^b_{n+1}, p^b_n)= [\, p^b_n q^b_n + L(q_n,q_{n+1})]
\,|_{q^b_n=h^b}.\label{F3}
\end{equation}
Then we have
\begin{equation}
{\partial F_3( q^a_{n+1},p^a_n) \over
\partial p^b_n} = q^b_n=h^b( q^a_{n+1}, p^a_n),
\end{equation}
and
\begin{equation}
{\partial F_3( q^a_{n+1},p^a_n) \over
\partial{ q^b_{n+1}}} = \left. {\partial L( q^a_{n}, q^a_{n+1}) \over
\partial{ q^b_{n+1}}} \right|_{q_n=h}=p^b_{n+1}.
\end{equation}

Notice that in the last equality there are also contributions coming
from the dependence on $q_{n+1}$ of the level $n$ variables $q_n$, but
these contributions cancel because of the definition of the canonical
momenta. The information about the momenta is completely encoded in
the evolution equations $q^a_{n+1}=h^a$ and the constraints.  As the
first equation is equivalent to (\ref{fun1}) one ends up recovering
the fundamental evolution equations as a canonical transformation
generated by $F_3$. Furthermore,
\begin{equation}
\left|{\partial F_3( q^a_{n+1},p^a_n) \over {\partial p^b_n
\partial q^a_{n+1}}}\right|= \left|{\partial q^b_{n} \over
\partial q^a_{n+1}}\right| \neq 0
\end{equation}
due to the fact that  we have assumed that $\D\left|{\partial
q^b_{n+1} \over
\partial q^a_{n}}\right| \neq 0$, and consequently $F_3$ is a
non singular generating function and therefore the resulting 
canonical transformation preserves the Poisson bracket
structure.

{\em Case II)} The equation(\ref{fun2}) is invertible for $p^a_n$:
that is $\D\left|{\partial p^a_{n+1} \over \partial p^b_n}\right|
\neq 0$ and therefore one can write,
\begin{equation}
p^a_n=g^a( q^b_n, p^b_{n+1}).\label{inv2}
\end{equation}

Notice that, under these hypotheses, there are no
pseudo-constraints of the form $G(q^a_{n},p^a_{n+1})=0$. By
substituting (\ref{inv2}) in (\ref{fun1}), one gets
\begin{equation}
q^a_{n+1}=k^a( q^b_n, p^b_{n+1}),\label{ev1}
\end{equation}
which allows to introduce a type 2 generating function
\begin{equation}
F_2( q^b_{n},  p^b_{n+1})= [\, p^b_{n+1} q^b_{n+1} -
L(q_n,q_{n+1})] \, |_{q^b_{n+1}=k^b}\label{F2}
\end{equation}

One can now easily check that this generating function reproduces
the evolution equations (\ref{fun1}) and (\ref{fun2}) and defines
a non singular canonical transformation that preserves the Poisson
Brackets in the evolution.

{\em Case III)} Even when the system has pseudo-constraints of the
form $G(q^a_{n+1}, p^a_n)=0$ and $F(q^a_{n}, p^a_{n+1})=0$ one may
be able to find a canonical transformation provided that the
system does not have pseudo-constraints of the form $F( p^a_n,
p^a_{n+1})=0$

In fact, by using Eq.(\ref{fun2}) one can invert for
\begin{equation}
q^a_{n}=l^a( p^b_n,  p^b_{n+1}),
\end{equation}
and substituting in (\ref{fun1}) get
\begin{equation}
q^a_{n+1}=m^a(p^b_n,  p^b_{n+1}).
\end{equation}

A generating function of type 4 that does the same job that the
two previous ones may now be introduced,
\begin{equation}
F_4(p^b_n,  p^b_{n+1})= \left. \rule[1ex]{0ex}{2ex}[-p^b_{n+1}
q^b_{n+1} + p^b_n q^b_n+ L(q_n,q_{n+1})]
\right|_{q^b_{n}=l^b}^{q^b_{n+1}=m^b}\label{F4}
\end{equation}
All of the discrete systems that have been treated up to now in
the literature may be analyzed by following this canonical procedure,
allowing to preserve the constraints and the Poisson bracket
structure. Later on we will show an example of a system of this type
in order to analyze how this procedure works in a concrete case.

It should be noted
that there may exist mechanical systems that do not fall into any
of the above classifications. For instance, a system could have
pseudo-constraints of all the types listed above. In such cases
one will need to develop further techniques to treat them.
For instance one could introduce canonical transformations of a
given type for some of the variables and of a different type for other
variables. This would require further study and it does not appear
necessary for the systems that have been analyzed up to present.

\section{Classification of the constraints}

At this point we have a set of constraints primary, secondary,
tertiary, etc, of the form ${\chi}_Z(q^a,p^a)=0$ with $Z=1...A$ with
$A$ the total number of independent constraints, that are preserved
under the evolution given by the equations(\ref{fun1},\ref{fun2})
provided part or all the arbitrary functions $V$ are conveniently
fixed.

As in the continuum case it is convenient to introduce the notion of
first and second class constraints, in order to quantize the theory. A
constraint is of first class if it commutes with all the constraints,
if that is not the case it is of second class.  As in the continuum
case one can define first class functions of the canonical variables
$f(q,p)$ that are not necessarily constraints. Such a function will be
first class if it commutes with all the constraints.  Second class
constraints may be imposed strongly by introducing Dirac brackets. As
the evolution equations preserve the Poisson structure, they will
preserve de Dirac structure because Dirac brackets are defined in
terms of Poisson brackets.  One ends up with a theory with a set of
evolution equations that preserve the symplectic structure of the
system, and therefore may be quantized by describing the evolution in
terms of unitary operators.

In the discrete case there is not a straightforward relation between
the number of first and second class constraints and the number of
phase space degrees of freedom.  This is due to the fact that now the
evolution of the constraints is not directly related with their
Poisson brackets with a total Hamiltonian. Thus, the fact that a
constraint does not commute with others is not easily related with the
determination of an arbitrary function. It is still very easy to
determine the number of phase space degrees of freedom.  In fact this
number is given by two times the number of configuration variables
minus the total number of constraints minus the number of arbitrary
functions $v$.

\section{An example of constrained system with second
class constraints}

To illustrate the techniques elaborated above, we would like to
discuss a model that is simple, yet addresses in a non-trivial way the
main points we discussed.  This example had been treated using ad-hoc
techniques in \cite{greece}. The model consists of a parameterized free
particle in a two dimensional space-time under the influence of a
linear potential. The discrete Lagrangian is given by,
\begin{equation}
L_n\equiv L(q_n^a,\pi_n^a,N_n,q_{n+1}^a,\pi_{n+1}^a,N_{n+1}) =
\pi_n^a(q_{n+1}^a-q^a_n)-N_n[\pi_n^0+
\frac{1}{2}(\pi_n^1)^2+\alpha q_n^1] \label{la}.
\end{equation}
We have chosen a first order formulation for the particle. However,
this Lagrangian is of the type we considered in this paper, one 
simply needs to consider all variables, $q^a,\pi^a,N$ as configuration
variables. The system is clearly  singular since the $\pi's$ and
$N$ only appear at level $n$ (or in the continuum Lagrangian, their
time derivatives are absent). When considered as a Type I
generating function, the above Lagrangian leads to the equations
\begin{eqnarray}
\p{p}{\pi}{a}{n+1} &=& \frac{\partial L_n}{\partial \pi^a_{n+1}}
=0, \label{evol11}
\\
\p{p}{q}{a}{n+1} &=& \frac{\partial L_n}{\partial q^a_{n+1}}
=\pi_n^a, \label{evol12}
\\
\p{p}{N}{}{n+1} &=& \frac{\partial L_n}{\partial N_{n+1}} =0,
\label{evol3}
\end{eqnarray}
and
\begin{eqnarray}
\p{p}{\pi}{a}{n} &=& -\frac{\partial L_n}{\partial \pi^a_n}
=-(q_{n+1}^a-q_n^a)+ \pi_n^1 N_n \delta^a_1+ N_n \delta^a_0,
\label{evo21}
\\
\p{p}{q}{a}{n} &=& -\frac{\partial L_n}{\partial q^a_n}
=\pi_n^a+\delta^a_1\alpha N_n, \label{evo22}
\\
\p{p}{N}{}{n} &=& -\frac{\partial L_n}{\partial N_n}
=\pi_n^0+\frac{1}{2}(\pi_n^1)^2+\alpha q_n^1 \label{evo23}.
\end{eqnarray}

One can easily recognize that the system has six constraints: three
at the $n+1$ level, and three at the $n$ level. They are:
\begin{eqnarray}
\psi_1^a &\equiv& \p{p}{\pi}{a}{n+1}=0, \label{const11}\\
\psi_2 &\equiv& \p{p}{N}{}{n+1}=0,\label{const12}
\\
\Phi_1^a &\equiv& \p{p}{q}{a}{n}- (\pi_n^a+\delta^a_1\alpha N_n),
\label{const01}\\
\Phi_2 &\equiv&
\p{p}{N}{}{n}-[\pi_n^0+\frac{1}{2}(\pi_n^1)^2+\alpha q_n^1].
\label{const02}
\end{eqnarray}
Therefore the evolution depends on three arbitrary functions
$\p{V}{N}{}{n},\p{V}{\pi}{a}{n}$,
\begin{eqnarray}
q_{n+1}^a &=& q_n^a + \pi_n^1 N_n \delta^a_1+ N_n \delta^a_0 
-\p{p}{\pi}{a}{n}\label{evolfun21}\\
\pi_{n+1}^a &=&\pi_n^a+\p{V}{\pi}{a}{n}\label{evolfun22}\\
{N}_{n+1} &=& {N}_{n} +\p{V}{N}{}{n}.\label{evolfun23}
\end{eqnarray}

The preservation of the $\psi$ constraints from level $n$ to level
$n+1$ is automatically ensured from (\ref{cons2}). Now we impose the
preservation of the $\Phi$ constraints upon evolution. Let us
begin with ${\Phi_1^0}$:
\begin{equation}
{\Phi_1^0}_{n+1} \equiv \p{p}{q}{0}{n+1}- \pi_{n+1}^0
=\p{p}{q}{0}{n}- \pi_{n}^0-\p{V}{\pi}{0}{n}=0,
\end{equation}
which taking into account the constraint ${\Phi_1^0}_{n}$
implies $\p{V}{\pi}{0}{n}=0$.

For the ${\Phi_2}$ one gets the equation
\begin{equation}
{\Phi_2}_{n+1} = \pi^0_n + \alpha(q^1_n+\pi^1_n N_n) +
(\pi^1_n+{V^\pi_n}^1)^2/2=0,
\end{equation}
that taking into account the constraint ${\Phi_2}_{n}$
implies that 
\begin{equation}
\p{V}{\pi}{1}{n}=-\pi^1_n + \epsilon
\sqrt{(\pi^1_n)^2-2\pi^1_nN_n\alpha}
\end{equation}
where $\epsilon=\pm 1$.

Finally we have 
\begin{equation}
{\Phi_1^1}_{n+1} =
\p{p}{q}{1}{n}-N_n\alpha-\pi^1_n-\p{V}{\pi}{1}{n}-\alpha(N_n+\p{V}{N}{}{n})
=0,
\end{equation}
that after imposing the constraint at level $n$ leads to
\begin{equation}
\p{V}{N}{}{n}=-\frac{1}{\alpha}\p{V}{\pi}{1}{n} -N_n.
\end{equation}

Thus, the evolution equations for the configuration variables are
\begin{eqnarray}
q_{n+1}^a &=& q_n^a + \pi_n^1 N_n \delta^a_1+ N_n \delta^a_0
\label{evolfunp21} \\ \pi_{n+1}^0 &=&\pi_n^0,\label{evolfunp220} \\
\pi_{n+1}^1 &=&\epsilon
\sqrt{(\pi^1_n)^2-2\pi^1_nN_n\alpha},\label{evolfunp221} \\ {N}_{n+1}
&=& \frac{1}{\alpha}[\pi_{n}^1 - \epsilon
\sqrt{(\pi^1_n)^2-2\pi^1_nN_n\alpha}].\label{evolfunp23}
\end{eqnarray}

We are now ready to define an invertible canonical
transformation with the help of a type 3 generating function.
Notice that these evolution equations are invertible for $q^a_n$,
and therefore we are in the case I. The inverse is given by a set
of equations of the form $q_n(q_{n+1})$, explicitly given by,
\begin{eqnarray}
q_{n}^1 &=& q_{n+1}^1 - \pi_{n+1}^1 N_{n+1} - \frac{\alpha}{2}
(N_{n+1})^2, \label{evolfunpa211} 
\\ q_{n}^0 &=& q_{n+1}^0-
\frac{\pi_{n+1}^1N_{n+1} + \D\frac{\alpha}{2} (N_{n+1})^2}
{\pi_{n+1}^1 + \alpha N_{n+1}},
\label{evolfunpa212}
\\
\pi_{n}^0 &=& \pi_{n+1}^0,\label{evolfunpa220}
\\
\pi_{n}^1 &=& \pi_{n+1}^1+\alpha N_{n+1},\label{evolfunpa221}
\\
N_n &=& \frac{\pi_{n+1}^1N_{n+1}+ \D\frac{\alpha}{2}
(N_{n+1})^2}{\pi_{n+1}^1+\alpha N_{n+1}}. \label{evolfunpa23}
\end{eqnarray}

Recalling that $F_3$ is given by equation (\ref{F3}) we obtain,
\begin{eqnarray}
F_3&=& \p{p}{\pi}{0}{n} \pi^0_{n+1} + \p{p}{\pi}{1}{n}( \alpha
N_{n+1} + \pi^1_{n+1} ) +  \p{p}{q}{0}{n} q^0_{n+1}
+\frac{1}{2}N_{n+1} \left[ \alpha^2 N^2_{n+1}+ 3 \alpha N_{n+1}
\pi^1_{n+1}+2(\pi^1_{n+1})^2 \right] \nonumber
\\
& &- \frac{N_{n+1}(\alpha N_{n+1} + 2\pi^1_{n+1})}{2(\alpha
N_{n+1} + \pi^1_{n+1})} \left[ \p{p}{q}{0}{n} - \p{p}{N}{}{n} +
\frac{1}{2}(\pi^1_{n+1})^2 + \alpha  q^1_{n+1} \right] +
\p{p}{q}{1}{n} \left(q^1_{n+1} - \frac{\alpha}{2} N_{n+1}^2 -
N_{n+1}\pi^1_{n+1} \right).
\end{eqnarray}

One can check that Eqs. (\ref{evolfunpa211}) to
(\ref{evolfunpa23}) are easily recovered by taking the partial
derivative with respect to $p_n^a$. By differentiating with respect to
$q_{n+1}^a$ one gets
\begin{eqnarray}
\p{p}{q}{0}{n+1} &=&\p{p}{q}{0}{n}, \label{evofin3}
\\
\p{p}{q}{1}{n+1} &=&\p{p}{q}{1}{n} - \frac{\alpha}{2} A_n N_{n+1}
(\alpha N_{n+1} + 2 \pi^1_{n+1}),  \label{evofin4}
\\
\p{p}{\pi}{0}{n+1} &=&\p{p}{\pi}{0}{n}
\\
\p{p}{\pi}{1}{n+1} &=&\p{p}{\pi}{1}{n} +\frac{N_{n+1}}{2} \left[
3\alpha N_{n+1} - 2 \p{p}{q}{1}{n} + 4 \pi^1_{n+1} + 2 B_n A_n
-A_n(\pi^1_{n+1}+B_n A_n)(\alpha N_{n+1} + 2 \pi^1_{n+1})\right],
\\
\p{p}{N}{}{n+1} &=&\frac{3}{2} \alpha^2 N_{n+1}^2 + \alpha
\p{p}{\pi}{1}{n} - \p{p}{q}{1}{n}\pi^1_{n+1} + (\pi^1_{n+1})^2 +
B_n \nonumber
\\
&&+ \alpha N_{n+1}\left[-\p{p}{q}{1}{n} + 3\pi^1_{n+1} -
\frac{1}{2} B_n A_n^2 (\alpha N_{n+1} + 2 \pi^1_{n+1})
\right],\label{ej-p2}
\end{eqnarray}
where we have introduced
\begin{eqnarray}
A_n &\equiv& (\alpha N_{n+1} +  \pi^1_{n+1})^{-1},
\\
B_n &\equiv& \p{p}{N}{}{n} - \p{p}{q}{0}{n} -
\frac{1}{2}(\pi^1_{n+1})^2 - \alpha  q^1_{n+1}.
\end{eqnarray}

By substituting in Eqs.(\ref{evofin3}) to (\ref{ej-p2}) the variables
$q_{n+1}$ and using the level $n$ constraints one gets Eqs.
(\ref{evol11}) to (\ref {evol3}), which is a canonical transformation
that reproduces, on the constraint surface, the evolution
equations of the discrete particle.

What remains to be done is to identify the second class
constraints and impose them strongly. The complete set of six
constraints of this model $\psi$ and $\Phi$ are second class and
allow to solve for $\pi^a$ and $N$ and eliminate completely  these
variables and their complex conjugates $P_\pi$ and $P_N$.

One can proceed in two different ways. The first alternative is to
start by observing that $P_\pi$ and $P_N$ vanish strongly, and
then to solve for $\pi$ and $N$ in terms of the $n+1$ level
variables, leading to
\begin{eqnarray}
\pi^0_n &=& \p{p}{q}{0}{n+1},
\\
\pi^1_n &=& \p{p}{q}{1}{n+1},
\\
N_n &=& \frac{C_{n+1}}{\alpha \p{p}{q}{1}{n+1}},
\end{eqnarray}
where $C_{n+1}=\p{p}{q}{0}{n+1} + (\p{p}{q}{1}{n+1})^2/2
+\alpha q^1_{n+1}$. The relevant evolution equations are obtained
from (\ref{evolfunp21},\ref{evofin3},\ref{evofin4}),and are given
by
\begin{eqnarray}
q^0_{n+1} &=& q^0_n + N_n =  q^0_n + \frac{C_{n+1}}{\alpha
\p{p}{q}{1}{n+1}},
\\
q^1_{n+1} &=& q^1_n + N_n \pi^1_n = q^1_n + \frac{C_{n+1}}{\alpha},
\\
\p{p}{q}{0}{n} &=& \p{p}{q}{0}{n+1},
\\
\p{p}{q}{1}{n} &=& \p{p}{q}{1}{n+1} + \alpha N_n =
\p{p}{q}{1}{n+1}+ \frac{C_{n+1}}{\p{p}{q}{1}{n+1}},
\end{eqnarray}
and we recover the evolution equations obtained in
\cite{greece}. 

The second alternative consists in solving for $\pi$ and $N$ in
terms of the $n$ level variables, leading to
\begin{eqnarray}
\pi^0_n &=& \p{p}{q}{0}{n},
\\
\pi^1_n &=& \epsilon \sqrt{-2 (\p{p}{q}{0}{n} +\alpha q^1_n)},
\\
N_n &=& \frac{1}{\alpha} \left( \p{p}{q}{1}{n} -\epsilon \sqrt{
-2(\p{p}{q}{0}{n} +\alpha q^1_n)} \right),
\end{eqnarray}
and from here, computing the evolution equations by
using (\ref{evolfunp21},\ref{evofin3},\ref{evofin4}). The two
methods yield evolution schemes of different functional form
since one propagates ``forward'' in time and the other ``backward''.
The inequivalence in the functional form stems from the fact that
the discretization of the time derivatives chosen in the Lagrangian
is not centered. It should be emphasized that if one starts from
given initial data and propagates forwards with the first system of
equations and then backwards using the second, one will return to
the same initial data.

Notice that we have 6 second class constraints, and the initial
number of phase space degrees of freedom was 10.
By noticing that there are no arbitrary functions left,
one is left with 4 degrees of freedom on the constraint surface.
The continuum model had two degrees of freedom.

The procedure we have followed here is completely general and may be
simplified when one is treating specific cases. For instance, as it
happens in the continuum theory \cite{teit} it is sometimes possible to
implement the canonical analysis by first solving the constraints for
the unphysical degrees of freedom $N,\pi,p^N,p^{\pi}$ and then
introducing a generating functional on the physical degrees of freedom
by following the procedure of the previous sections. In this
particular case it is easy to show that, for $p^{q^1}\neq 0$,
$\D\left|{\partial q^a_{n+1} \over \partial q^b_n}\right|\neq 0$,
where $q^a \equiv (q^1,q^0)$, and thus it is possible to construct an
$F_3$ generating functional. In all the models treated up to now in
the literature the unphysical degrees of freedom were eliminated
before obtaining the canonical transformation for the evolution of the
physical degrees of freedom. In order to keep the analysis general in
a simple model, here we have kept all the variables involved in this
approach.

\section{Treatment in terms of Type II generating functions}

Up to now, we have taken as the starting point a singular Type I
generating function given by $F_1= -L( q^a_n, q^a_{n+1})$. It is
interesting to analyze how singular systems may be described in
terms of other types of generating functions.

Let us assume that neither condition (\ref{det}) nor condition
(\ref{det0}) are fulfilled by the discrete system and, therefore, we
are in the singular case. Then, it may be immediately seen, by using
(\ref{P1}),and taking into account that the determinant that appears in
(\ref{det}) now vanishes, that singular systems with first order
Lagrangians in the continuum lead to the presence of pseudo-constraints
$\phi_{\alpha}(q_n,p_{n+1})=0$, in the theory.  Recall that these
pseudo-constraints are similar to the constraints that arise in the
continuum case, but mixing configuration variables at one level with
momenta in the next level. As in the continuum case, one can then
introduce the {\it pseudo}-constraint surface $S_{\phi}$ which we
shall assume has well defined functions,
\begin{equation}
\widetilde{\nabla}\phi_{\alpha} = \left(
\ldots\frac{\partial\phi}{\partial q^a_n}\ldots,
\ldots\frac{\partial\phi}{\partial p^a_{n+1}}\ldots \right)\,,
\end{equation}
where a vector $\tilde{\tau}$ in the tangent space of $S_\phi$,
$T_{(q,p)} S_\phi$, is such that
\begin{equation}\label{EspTangente}
\tilde{\tau}\cdot\widetilde{\nabla}\phi_{\alpha}=0 \;\; \forall
\alpha.
\end{equation}

Let us introduce now as before the type-2 Hamiltonian
(\ref{Ham2}),
\begin{equation}
H_2 \equiv  \sum_a p^a_{n+1} (q^a_{n+1} - q^a_n) - L(q_n,q_{n+1})\; ,.
\end{equation}

For which it is easy to see, using the pseudo-constraints, that it is
a function of $q_n,p_{n+1}$.  Let us consider now an infinitesimal
variation of $H_2$ along the pseudo-constraint surface,
\begin{eqnarray}
dH_2 &=& \sum_a \left[(q^a_{n+1} - q^a_n) dp^a_{n+1} +
\left(p^a_{n+1} - \frac{\partial L}{\partial q^a_{n+1}} \right)
dq^a_{n+1} - \left(p^a_{n+1} + \frac{\partial L}{\partial q^a_n}
\right) dq^a_n \right] \nonumber
\\
&=&\sum_a \left[(q^a_{n+1} - q^a_n) dp^a_{n+1}  - \left(p^a_{n+1}
+ \frac{\partial L}{\partial q^a_n} \right) dq^a_n
\right],\label{dH}
\end{eqnarray}
where in the last step we used (\ref{1}). $H_2$ is well
defined in $S_\phi$ but can be extended to the whole phase space
as it is done in the case of constraints in the continuum theory.
In order to obtain the canonical equation of motion we start from
the identity,
\begin{equation}
dH_2(q_{n},p_{n+1}) = \sum_a \left(\frac{\partial H_2}{\partial q^a_n}
dq^a_n +\frac{\partial H_2}{\partial p^a_{n+1}} dp^a_{n+1}
\right)\,,
\end{equation}
and evaluating it for an infinitesimal displacement $(dq,dp)$ in
$T_{(q,p)}S_\phi$, we can use (\ref{dH}) and the Lagrangian
equations of motion (\ref{2}) to obtain,
\begin{equation}
\sum_a \left[ \left(p^a_{n+1} - p^a_n + \frac{\partial H_2}{\partial
q^a_n} \right) dq^a_n + \left(q^a_n-q^a_{n+1} + \frac{\partial
H_2}{\partial p^a_{n+1}} \right) dp^a_{n+1} \right]=0.\,
\end{equation}

Since $(dq,dp)$ is an arbitrary tangent vector to $S_\phi$,
following (\ref{EspTangente}) we obtain that the coefficient
must be proportional to the gradient. Introducing therefore the
Lagrange multipliers $\lambda^a$ as the proportionality factors, we
will end with the set of equations,
\begin{eqnarray}
q^a_{n+1} &=& q^a_{n} +{\partial H_2 \over \partial p^a_{n+1}} +
\lambda^{\alpha}_n{\partial \phi_{\alpha} \over
\partial p^a_{n+1}},\label{h1}
\\
p^a_{n+1} &=& p^a_{n} -{\partial H_2 \over
\partial q^a_{n}} - \lambda^{\alpha}_n {\partial \phi_{\alpha} \over
\partial q^a_{n}}, \label{h2}
\\
\phi_{\alpha}(q_n,p_{n+1})&=& 0. \label{c}
\end{eqnarray}

This system is similar to the continuum set of equations with the
inclusion of pseudo-constraints, which have the same functional form
as the continuum constraints, but involve configuration variables at
level $n$ and momenta at level $n+1$, instead of $n$ level
variables. From this initial evolution equations one may follow a
procedure similar to the one developed in the previous section in
order to study the consistency of the (primary) constraints and
pseudo-constraints.

\section{Discussion and conclusions}

We have provided a canonical procedure for the introduction of a
preserved symplectic structure in discrete constrained systems.
The analogy with Dirac's procedure in the continuum is quite
remarkable. It is possible to define a notion of discrete
evolution that weakly preserves constraints and Poisson brackets.
The distinction between first and second class constraints is
still useful and when second class constraints are imposed
strongly the resulting Dirac brackets are preserved. 

A feature of the discretized theories is that they may have a smaller
number of first class constraints, and consequently more degrees of
freedom than the continuum counterparts. The extra degrees of freedom
come from the fact that the discrete theories may not necessarily have
the same symmetries as the continuum theories. For instance, in the
case of homogeneous cosmologies studied in \cite{cosmo} the extra pair
of phase space degrees of freedom are associated with the fact that in
the discrete theory different choices of refinements in the
discretization in time correspond to different solutions in the
discrete theory that nevertheless approximate the same solution in the
continuum theory.

This is only a first step for a complete understanding of the dynamics
of discrete gauge systems. The relation between the number of
constraints of first and second class and the number of degrees of
freedom, and the connection between the first class constraints and
the gauge invariance of the discrete dynamical system need to be
further studied. Moreover, as discussed in the body of the article, if
one wishes to consider more pathological systems than the ones
considered here, more elaborate canonical transformations may need
to be introduced.

The issue of the continuum limit is well understood in the non
singular case, where there is an external step parameter that
controls the approximation. However, it needs further study in the
case of singular systems, particularly in the case of totally
constrained systems where the step of the approximation is encoded
in the additional degrees of freedom of the discrete theory. This
issue has been studied in several models\cite{cosmo,greece} but a
complete characterization of the possible behaviors is still
lacking. A similar comment applies to the role of spatial 
discretizations when one is considering lattice field theories.

\section{Acknowledgments}
This paper was motivated by a series of questions by Karel Kucha\v{r}
about our previous work.  This work was supported by grant
NSF-PHY0244335 and funds from the Horace Hearne Jr. Laboratory for
Theoretical Physics
and the Abdus Salam International Center for Theoretical Physics.

\end{document}